\documentclass[sigconf,nonacm]{acmart}
\usepackage{enumitem}
\usepackage{soul}
\usepackage{makecell}
\usepackage{multirow}
\usepackage{fontawesome5}
\usepackage{hyperref}
\usepackage{booktabs}
\usepackage{subcaption}
\usepackage{microtype}
\usepackage{tabularx}
\AtBeginDocument{}

\title{Customizing an LLM for Enterprise Software Engineering}

\begin{document}

\settopmatter{authorsperrow=4}
\author{Aditya Kini}
\email{akini@google.com}
\affiliation{%
  \institution{Google}
\country{}}

\author{Satish Chandra}
\authornote{Work done while at Google.}
\email{schandra@acm.org}
\affiliation{%
  \institution{Meta}
\country{}}

\author{Milad Hashemi}
\email{miladh@google.com}
\affiliation{%
  \institution{Google}
\country{}}

\author{Saksham Thakur}
\email{sakshamthakur@google.com}
\affiliation{%
  \institution{Google}
\country{}}

\author{Aditya Pandey}
\email{adityapandey@google.com}
\affiliation{%
  \institution{Google}
\country{}}

\author{Vincent Nguyen}
\email{vincentdnguyen@google.com}
\affiliation{%
  \institution{Google}
\country{}}

\author{Marc Brockschmidt}
\email{mmjb@google.com}
\affiliation{%
  \institution{Google}
\country{}}

\author{Franjo Ivan\v{c}i\'{c}}
\email{ivancic@google.com}
\affiliation{%
  \institution{Google}
\country{}}

\author{Danny Tarlow}
\email{dtarlow@google.com}
\affiliation{%
  \institution{Google}
\country{}}

\author{Parthasarathy Ranganathan}
\email{parthas@google.com}
\affiliation{%
  \institution{Google}
\country{}}

\author{Petros Maniatis}
\email{maniatis@google.com}
\affiliation{%
  \institution{Google}
\country{}}

\author{Ahmed Omran}
\email{ahmedomran@google.com}
\affiliation{%
  \institution{Google}
\country{}}

\author{Zaheer Abbas}
\email{zaheersm@google.com}
\affiliation{%
  \institution{Google}
\country{}}

\author{Anita Gergely}
\email{agergely@google.com}
\affiliation{%
  \institution{Google}
\country{}}

\author{Martin Sevenich}
\email{msevenich@google.com}
\affiliation{%
  \institution{Google}
\country{}}

\author{Gufeng Zhang}
\email{gufengzhang@google.com}
\affiliation{%
  \institution{Google}
\country{}}

\author{Amy Hua}
\email{amyhhua@google.com}
\affiliation{%
  \institution{Google}
\country{}}

\author{Alexander Frömmgen}
\email{froemmgen@google.com}
\affiliation{%
  \institution{Google}
\country{}}

\renewcommand{\shortauthors}{Kini et al.}

\begin{abstract}
  Enterprise software development is a continuous evolutionary process, characterized by incremental additions, architectural revisions, production deployments and rigorous maintenance. These activities generate valuable data that modern LLMs could be finetuned on, to unlock additional tool possibilities for enterprise software engineering.  While frontier LLMs are already very capable, this form of customization offers a compelling path for enterprise-specific optimization.

  We introduce
  \textit{Gemini for Google (GfG)}, an adaptation of Gemini specialized for Google's internal software engineering ecosystem. This paper details the model's end-to-end development, from curating a trillion-token proprietary dataset to implementing a mid-training strategy that mitigates catastrophic forgetting. In a
  large-scale blind A/B study across 29,000 developers, Gemini for Google significantly outperformed baselines: reducing the mean number of iterations per turn by 23\%, and increasing code survival rates by about 17\%. Beyond metrics, we provide a comprehensive blueprint for enterprise model adaptation, covering: (1)~the extraction of high-value signals from software engineering data,
  (2)~data preparation strategies, (3)~full-stack model tuning (continued pre-training and post-training), and (4)~the deployment of downstream applications. We believe this methodology offers a replicable path for other organizations to unlock the full potential of their internal engineering data.
\end{abstract}


\maketitle
\section{Introduction}
Building, maintaining, and evolving the software solutions that power large organizations requires an ecosystem of engineering activities: from design and documentation, to code generation, code verification, code review, rollouts, capacity management, and production monitoring. Each of these tasks requires specialized, often enterprise-centric and customized human expertise, and often generates artifacts, tools, and documents that are unique to a particular company or a particular codebase.

Many recent products and research projects are centered around using large language models (LLMs) to accelerate software development, but often focusing mostly on high-visibility use-cases like code generation. This shift is already operating at a massive scale, as publicly disclosed figures indicate that over 75\% of code at Google is now written or drafted by AI \cite{pichai_cloud_next_2026}. Despite this prevalence, standard evaluation benchmarks, such as HumanEval, primarily assess a model's ability to generate greenfield solutions for algorithmic challenges. However, this focus on algorithmic generation creates a misalignment with industrial requirements. In practice, the majority of enterprise engineering effort is not dedicated to self-contained or context-independent code generation, but to the maintenance and evolution of existing systems: large-scale refactoring, efficiency optimization, and program comprehension within legacy environments.

To effectively assist in these high-value maintenance tasks, LLMs must reason over the specific semantics of an organization's proprietary ecosystem. At Google, this comprises a monolithic codebase replete with custom coding paradigms, deep integration of proprietary libraries, and high-velocity development patterns that create a continuously shifting context for any AI assistant. General-purpose frontier models possess strong reasoning capabilities but lack exposure to the unique data distributions---internal libraries, architectural patterns, and domain-specific languages—that are absent from public training sets.

There are multiple strategies for making this enterprise data visible to, or at least influence, the inference carried out by a foundation LLM. One common approach used by many companies is to leverage retrieval mechanisms and prompt engineering (RAG) on top of generic LLMs without any specialization. The results of retrieval augmentation are quite mixed and depend on the quality of the retrieval and the manner in which the context is populated. Another method is to use an adaptation technique such as LoRA; however (owing to low rank) they lack the representational capacity required for a large-scale domain adaptation of a trillion-token corpus. Finally, organizations may choose to train smaller language models from scratch using enterprise data, a strategy that prioritizes complete control over the model architecture and training lifecycle.

In this work, we present a complementary methodology: the direct specialization of foundation models via mid-training intervention and full-parameter fine-tuning. Unlike training from scratch, this approach branches off from a foundational checkpoint, preserving general reasoning capabilities while injecting deep domain knowledge. We propose a comprehensive blueprint for enterprise adaptation, detailing the extraction of high-value signals from software engineering data, data preparation strategies, and full-stack model tuning. While our implementation utilizes the Gemini foundation model, the core contribution of this paper is a replicable framework for data curation and training dynamics. This methodology allows organizations to apply these techniques to any available high-capability models, creating enterprise experts capable of addressing the full spectrum of development tasks.

For brevity's sake, a direct quantitative comparison to retrieval-based or scratch-trained approaches is outside the scope of this article. Training a model from scratch was not pursued due to the prohibitive computational cost and time constraints relative to the efficiency of our intervention strategy. Additionally, the evaluation of open-weight architectures was constrained by incompatibilities with our proprietary distributed training infrastructure and strict data governance protocols regarding the ingestion of internal code.

We describe our process for curating the dataset, and provide three case studies on how the resulting model is used as a ``first-party software development assistance specialist'', reducing the human cost of developing and maintaining an enterprise-scale monorepo consisting of billions of lines of code, and improving the efficiency of Google's planetary-scale data centers.

The customized model, \textit{GfG}, has been in use at Google to serve IDE features~\cite{tabachnyk2026achievingproductivitygainsaibased}, to carry out
large-scale migrations~\cite{nikolov2025googleusingaiinternal} and performance-improving code refactorings~\cite{lin2025ecollmdrivenefficientcode}, and in other important use cases.  Evaluations in these contexts have been presented in the respective articles;  we recap these briefly in \S\ref{sec:results}. Further, in this article, we present online evaluations with developer productivity at Google in general as well as offline evaluations with selected coding benchmarks.

\textbf{Contributions.} This paper makes the following contributions:
\begin{itemize}[label=$\square$, topsep=0pt, itemsep=0pt]
  \item \textbf{A Framework for Enterprise Adaptation:} We provide a blueprint for building a specialist foundation model from internal engineering data covering data curation, training, and deployment.
  \item \textbf{Mid-Training Intervention Strategy:} We employ a mid-training intervention to bridge the gap between general pre-training and post-training via a \textit{torso-patch} to adapt model weights for Google's internal software engineering ecosystem.
  \item \textbf{Enterprise-Scale Evaluation:} We demonstrate the effectiveness of this approach in a blind A/B study with over 29,000 Google developers showing that Gemini for Google (\textit{GfG}) significantly outperforms the baseline Gemini model by reducing conversational iterations while increasing code acceptance and survival rates.
\end{itemize}

{\bf Overview.} The rest of the paper is structured as follows. We next provide some background on our developer ecosystem in \S\ref{sec:background}, before we define the data curation methodology that addresses the core target capabilities demanded of a bespoke enterprise model in \S\ref{sec:methodology}. Next, we detail the mid-training intervention strategy used to specialize the model in \S\ref{sec:midtraining} and the subsequent supervised fine-tuning in \S\ref{sec:posttraining}. We then evaluate through online and offline results such as large-scale A/B study and case studies in \S\ref{sec:results}. We conclude with lessons learned and a discussion of future work in \S\ref{sec:lessons}.

\section{Background: Google's Engineering Ecosystem}
\label{sec:background}
Our study is situated within a hyperscale industrial software environment tailored to support global-scale services. Development is centralized within a single monolithic repository (``google3'') comprising billions of lines of code. This environment is characterized by a high degree of reliance on proprietary infrastructure rather than standard open-source equivalents:
\begin{itemize}
  \item \textbf{Core Libraries and Infrastructure}: Extensive use of internal libraries/frameworks and custom distributed storage (colossus ~\cite{HildebrandSerenyi2021}) and processing engines (e.g. F1~\cite{41344}, flume~\cite{35650}, Borg~\cite{43438}, Chubby~\cite{27897}).
  \item \textbf{Development Workflow}: Engineering activities are conducted via specialized internal tools, including the Cider IDE and Clients in the Cloud (CitC) for workspace management.
  \item \textbf{Quality Assurance And Review}: Code evolution is managed through a rigorous peer-review process via Critique ~\cite[Chapter 19]{50283}. Submissions--known internally as Changelists (CLs), similar to PRs or diffs--undergo automated static analysis, linting, and regression testing before and during the review process. Reviewers focus on correctness, maintainability, and test coverage, creating a collaborative dialogue that iteratively refines the codebase.

  \item \textbf{Observability}: A bespoke suite of production monitoring tools provides fine-grained logging, tracing, and efficiency analysis.
\end{itemize}

Crucially, this entire lifecycle—from IDE interactions to code review comments and production logs—generates a massive, structured corpus of software engineering data. Unlike primarily code-centric datasets, this internal data captures the full provenance of software evolution, providing the necessary signal to train models on the nuanced requirements of an enterprise stack ~\cite{maniatis_tarlow_2023}.

\section{Methodology}
\label{sec:methodology}
In order to develop an enterprise expert model, we first define the necessary core capabilities to support Google's internal development workflows. These requirements then inform our data curation strategy to ensure alignment with the downstream applications.

\subsection{Target Capabilities}
We focus on high-value areas to provide signficant value to Google's engineering workflows:

\begin{itemize}
  \item \textbf{Code Transformations / Automated Program Transformation~\cite{tabachnyk2026achievingproductivitygainsaibased}}: We envision an LLM that acts as a ``code transformation specialist'', capable of
    (a) automatically refactoring code, (b) improving performance, (c) migrating code to new APIs, and (d) fixing bugs. Given the size and complexity of Google's codebase, this capability offers the potential for improving our engineering velocity by automating some code maintenance tasks.

  \item \textbf{Internal Chatbots/Knowledge Discovery}: Navigating Google's internal documentation and APIs can be challenging given our scale. A Google-aware model has the ability to understand user queries about internal products, which is especially useful given that many internal systems have proprietary names.

  \item \textbf{Code Completion}: While smaller code-generation models were deployed at Google before this work, our intent was to push the boundary in this space using Gemini models: producing a code generation engine that understands larger contexts and anticipates developer intent, writing an ever more substantial fraction of Google's code.

\end{itemize}

There are several additional use cases that benefit from a custom model, e.g. assistance with debugging, but we focus on the above topics in this paper.

\subsection{Data Curation}
\label{sec:data_curation}

To satisfy the above requirements, we curated a corpus capturing diverse artifacts generated during software development. We categorize this dataset into six primary domains, each providing distinct supervision signals for the model.

\begin{enumerate}[leftmargin=*, nosep]
  \item \textbf{Change-Generation}: Google's rigorous code review process provides a broad set of data on code transformations. By including code review data and developer conversations, we provide the LLM with information on imperative code transformation - showing how code is improved, refactored, and optimized.

    One representative source of data in this category is the comment resolution dataset. Building this dataset involves extracting tuples of the initial code state, comment, and final code state:
    \begin{equation*}
      (C_\text{initial}, \text{comment}, C_\text{final})
    \end{equation*} where the difference between  $C_\text{final}$  and $C_\text{initial}$  represents the semantic edit required to resolve the reviewer's comment. As discussed in prior work\cite{52878}, this data enables the model to learn the nuances of design, complexity, and maintainability by modeling the {\em ``critique-and-refine''} dialogue inherent to high-quality software engineering.

    \begin{figure}
      \centering
      \includegraphics[width=\linewidth]{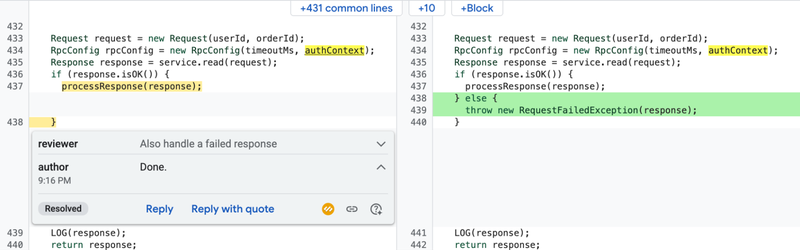}
      \caption{A sample interaction in the code review tool which was resolved by relevant code edits by the code author}
    \end{figure}

  \item \textbf{Knowledge}: To power an internal chatbot, we need to provide the LLM with a deep understanding of Google's internal knowledge base. To this end, we aggregate unstructured and semi-structured knowledge artifacts, such as internal technical documentation, API references, design documents and an internal Q\&A repository.

    \begin{figure}
      \centering
      \includegraphics[width=\linewidth]{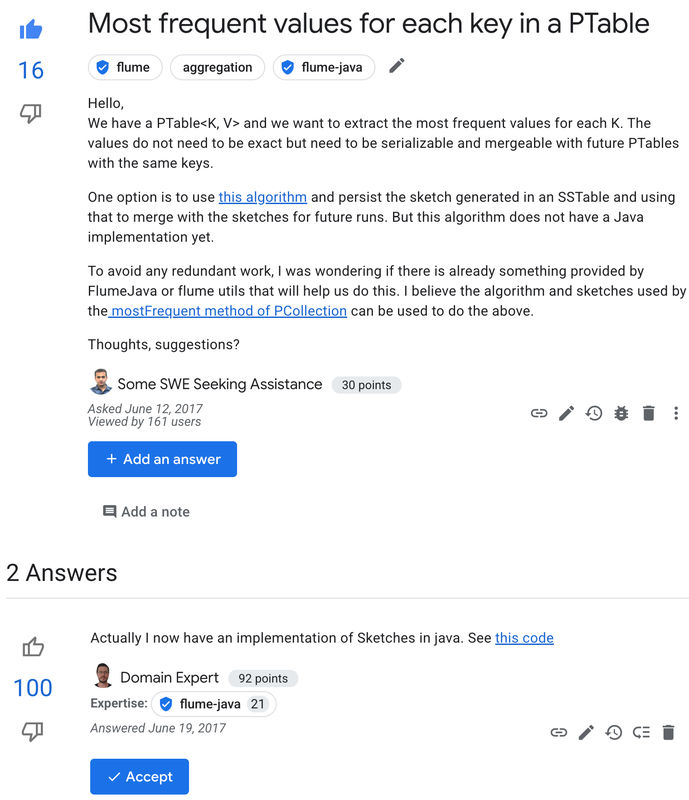}
      \caption{An example of a question asked by a SWE on Google's internal Q\&A tool. These discussions between developers and domain experts can be synthesized to form good knowledge category training data.}
    \end{figure}

  \item \textbf{Issues and Fixes}: We leverage Google's large internal datastore for capabilities like bug fixing, build errors fixing, test failures and other issues. By including data on these issues and their corresponding code fixes, we provide the LLM with valuable data on development context, how developers break down and solve problems, and how we diagnose production events that may arise.

    A key contribution in this domain is the Build Fixing dataset. By utilizing fine-grained IDE instrumentation, we reconstruct developer sessions as a sequential series of file snapshots. We identify discrete intervals [time\_initial ($t_i$), time\_final($t_j$)]  where the build state transitions from \textit{Failure} $\rightarrow$ \textit{Success}. The code difference $\Delta(t_i, t_j)$, contextualized by the corresponding error at $t_i$, provides a source of implicit supervision for automated program repair (APR), allowing us to extract millions of training examples without manual annotation~\cite{johnston2024safely}.

    \begin{figure}
      \centering
      \includegraphics[width=\linewidth]{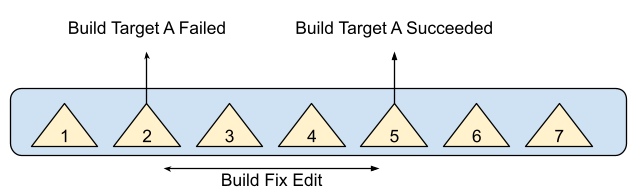}
      \caption{A developer's IDE editing session is visualized as a series of snapshots, numbered sequentially. Fine-grained logging within the IDE captures each editing action. In this example, a build failure occurred at snapshot 2, and the build was successful at snapshot 5. Consequently, the changes made between snapshot 2 and snapshot 5 represent the edits that resolved the build issue.
      }
    \end{figure}

    Other types of data in this category include test failure fixes, bug fixes, fixes for errors discovered by the wide range of static analyzers, and fixes for style guide violations.

  \item \textbf{Code Generation~\cite{tabachnyk2022ml}}: To support cases like code completion we include code written by Google developers, processed into multiple formats (e.g., FIM \cite{bavarian2022efficienttraininglanguagemodels}) and gran\-u\-lar\-i\-ties (single file, multi-file) to improve the model's ability to generate both short and long-form code completions.

  \item \textbf{Logs and Performance}: This category includes a subset of Google's production metadata and events - focusing on code efficiency and debugging traces. This kind of data provides context to the model to develop efficient and reliable software systems.  One notable source of data in this category is our repository of efficiency edits. This is helpful in bridging the gap between static code and dynamic execution. This corpus is derived from a history of optimization changes targeting production services. By identifying commits explicitly linked to latency reduction or CPU/memory optimization—and pairing them with the performance anti-patterns they resolved—we train the model to recognize and suggest efficiency improvements. This aligns the model's generation objectives with production health metrics~\cite{lin2025ecollmdrivenefficientcode}.
    We also incorporate sources like Dapper~\cite{dapper} — a system used for tracing and visualizing computations across Google's distributed computation stack—to capture the causal and temporal structure of these operations, providing critical visibility into runtime behaviors such as inter-service dependencies and latency bottlenecks. By training on these traces the model learns to reason about complex system interactions and diagnose execution-path inefficiencies that are often invisible to static analysis tools.

  \item \textbf{Activity Timelines}: Finally, we incorporate coarse-grained workflow data to capture the broader context of engineering tasks. This includes activity timelines that map tool usage, search queries, and documentation access patterns, providing the model with a holistic view of how engineers navigate the ecosystem to complete complex tasks~\cite{maniatis_tarlow_2023}.
\end{enumerate}

\subsection{Data Format}
While transformer-based, decoder-only architectures inherently rely on autoregressive (causal) modeling, where, for a continuous stream of tokens $x = (x_1, \dots, x_n)$, the likelihood $P(x) = \prod_{i} P(x_i \mid x_{<i})$ is maximized for a subsequent token. Applying this objective naively to heterogeneous software engineering tasks yields suboptimal results. To address this, we design our data such that the model can not just learn correct syntax, but also deduce and correctly predict code changes requested by a user through some natural-language instructions, using autoregressive decoding only. When training on the causal language modeling task, we apply masking of the input tokens to limit the loss calculation exclusively to the modified tokens.

In the following, we present some of the employed data representations:\footnote{The scope of this paper is limited to human-generated trajectories; the analysis of synthetic or agent-generated data falls outside the current research objectives.}

\begin{enumerate}[leftmargin=*, nosep]
  \item \textbf{Unstructured natural language and semi-structured operational data (e.g., technical documentation, production logs)}:  This data covers the Knowledge and Operational Semantics categories, where the objective is next-token prediction based on historical context:
    \begin{equation*}
      P(\text{Target}) = P(\text{Token } | \text{ Prefix, Prefix} + \text{Token} \in \text{Document})
    \end{equation*}
  \item \textbf{Instruction-Guided Code Editing (Code-to-Diff)}   Here, we provide the model with some context, including some original code and a user instruction on the desired changes ($\text{Code}_{\text{original}}$, $\text{Instruction}$). The model's target is to predict a code diff $\Delta$ that satisfies the instruction:
    \begin{equation}
      P(\text{Target}) = P(\Delta \mid \text{Code}_{\text{original}}, \text{Instruction}).
    \end{equation}
    One example for a commonly used diff format is the unified diff, shown in Fig.~\ref{fig:comment_resolution_diff}.

    A wide range of datasets from the Change Generation category mentioned above fit into this representation. Some examples (not exhaustive) are listed below.

    \begin{itemize}
      \item \textbf{CL Description to Change}: Derived from submitted changelists (CLs), where the ``Instruction'' is the CL description and the target is the submitted patch.

      \item \textbf{Code Comment Resolution}: Derived from peer-review interactions. The input is the code snippet plus the reviewer's comment; the target is the edit that resolved the comment~\cite{52878}.
        This code diff representation also extends well into the issues and fixes category of datasets, where the issue is described using its natural language description or by annotating the code with specific issues like compilation errors or test failures:
        \begin{equation}
          P(\text{Target}) = P(\Delta \mid \text{Code}_{\text{broken}}, \text{Error}_{\text{signal}}).
        \end{equation}
        One example from this category that fits into this representation is listed below.

    \end{itemize}

    \begin{figure}
      \centering
      \includegraphics[width=\linewidth]{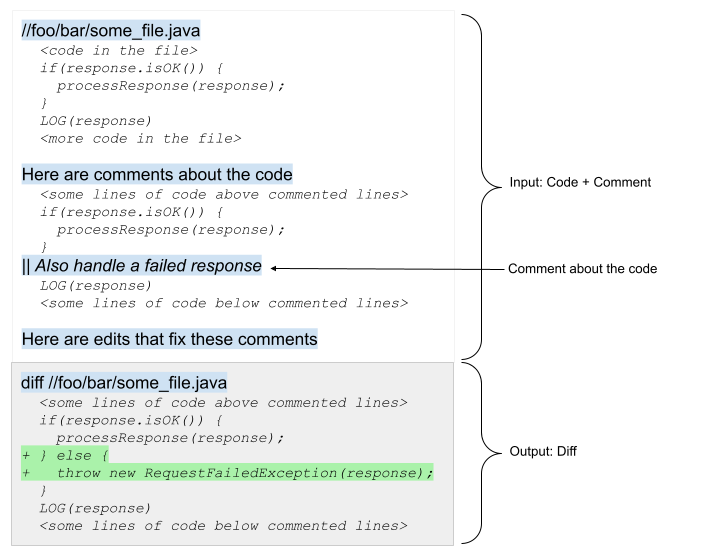}
      \caption{A comment resolution example.}
      \label{fig:comment_resolution_diff}
    \end{figure}

  \item \textbf{Automated Program Repair (APR)}:   We adapt the diff format for the Issues \& Fixes category. Here, the ``Instruction'' is structured as a compiler error or static analysis warning, and the target is the fix. By modeling compiler errors as ``automated review comments,'' we unify the representation of human and machine feedback, as described in~\cite{johnston2024safely}.

  \item \textbf{Context-Aware Infilling (FIM)}: This format is used to support cases like code completion within existing files. This “Code Generation” category of data uses FIM-based(Fill in the middle) formats ~\cite{bavarian2022efficienttraininglanguagemodels}, where the code section to be predicted is split into a suffix, middle and prefix with sentinel tokens used to delineate the different sub-sequences.

    This training method enhances the infilling ability of the models which have strong autoregressive generation abilities. By employing a distinct preprocessing technique, the method divides code into three segments: Prefix, Middle, and Suffix. These segments are then rearranged for training, allowing Middle tokens to rely on both Prefix and Suffix rather than solely Prefix, as typical in causal attention models. We also apply FIM to multi file examples where context across files also aids the model's ability to generate code. We leverage three types of formats for this dataset:

\begin{verbatim}
1. Untransformed: <filepath>\n<content>[eod]
2. PSM: <filepath>\n[pre]<prefix>[suf]<suffix>
[mid]<middle>[eod]
3. SPM2: <filepath>\n[pre][suf]<suffix>[mid]
<prefix><middle>[eod]
\end{verbatim}

    The corresponding data with multi-file examples is shown in Fig.~\ref{fig:fim}.

    \begin{figure}
      \centering
      \includegraphics[width=\linewidth]{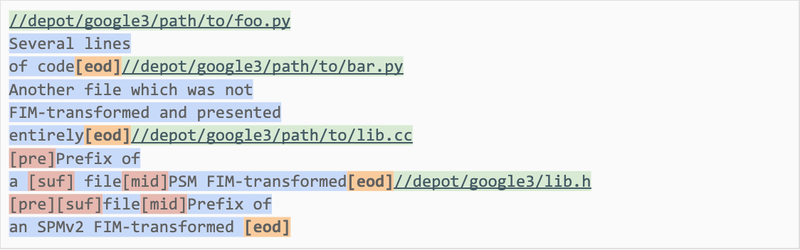}
      \caption{\label{fig:fim}A multi-file example.}
    \end{figure}

  \item \textbf{Multi-Turn Agentic Trajectories.}
    In order to capture the iterative nature of software development we include multi-step journeys datasets from extended periods of time. These data sources could be “bugs”, where developers compile sets of changelists to complete large tasks, or Activity Timelines, which show what actions developers took over shorter periods of time as they write code. To represent this data, we concatenate the text of the different issues or developer actions alongside the generated code in diff format, up to the size of the training window.

    A prominent example in this category is the Change Generation ReAct dataset. Leveraging the \textbf{ReAct} (Reasoning and Acting) framework~\cite{yao2023reactsynergizingreasoningacting}, we synthetically reconstruct the latent development trajectory for historical changelists. Rather than mapping a description directly to a final diff, we model the plausible sequence of intermediate actions -- such as code search queries, file navigation, and reasoning steps -- that a developer would undertake to implement the specified change.

\end{enumerate}

\section{Mid-Training Intervention}
\label{sec:midtraining}
Using the datasets above, we compile a trillion token-scale data mixture for auto-regressive training, similar to prior work~\cite{roziere2023codellama, lozhkov2024starcoder2, guo2024deepseekcoder, grattafiori2024llama3}. This is accomplished by intervening towards the end of pretraining, injecting our datasets, and continuing to train the model on the expanded corpus.

We train two distinct model variants to address different latency and capability requirements:
\begin{itemize}
  \item \textbf{Gemini-Flash based}: Optimized for low-latency and high-through\-put tasks such as real-time code completion.
  \item \textbf{Gemini-Pro based}: Optimized for complex reasoning tasks such as refactoring and architectural Q\&A.
\end{itemize}

Our final mixture consists of these categories of data mixed in with some breakdown that works best for our downstream applications. Table~\ref{tab:task_distribution} showcases a sample distribution for illustrative purposes only.

\begin{table}[ht]
  \centering
  \caption{Distribution of Software Engineering Tasks}
  \label{tab:task_distribution}
  \begin{tabular}{lr}
    \toprule
    \textbf{Task Category} & \textbf{Percentage (\%)} \\
    \midrule
    Code Generation        & 25.0 \\
    SWE Timelines          & 23.0 \\
    Issues and Fixes       & 22.0 \\
    CL Generation          & 20.0 \\
    Knowledge              & 5.0  \\
    Logs and Performance   & 5.0  \\
    \midrule
    \textbf{Total}         & \textbf{100.0} \\
    \bottomrule
  \end{tabular}
\end{table}

\subsection{Training-Time Evaluation}
During the mid-training intervention phase, we continuously assess intermediate checkpoints to validate convergence and ensure alignment with training objectives. These \textit{training-time} signals facilitate evaluation against real-world enterprise scenarios using a curated dataset derived from product logs—including IDE interactions and developer assistance features.

Our evaluation suite encompasses a diverse set of tasks:
\begin{enumerate}
  \item \textbf{Code Transformation}: Natural language-to-code tasks derived from logs of our internal IDE and manually curated for quality assurance.
  \item \textbf{Code Q\&A}: Conversational code queries featuring expert-verified responses.
  \item \textbf{Unit Test Generation}: Automated generation of test cases for code diffs.
  \item \textbf{Code Completion}: Analysis of model recommendations based on historical developer acceptance logs.
  \item \textbf{Code Migration}: Automated refactoring (e.g., int32 to int64 transitions).
  \item \textbf{Efficiency Refactoring}: Edits optimized for performance improvement.
  \item \textbf{Build Repair}: Automated resolution of build failures.
\end{enumerate}

We assess model performance using a multi-dimensional framework:

\begin{enumerate}
  \item \textbf{Static Analysis}: Code similarity metrics (e.g., fuzzy match, BLEU) against the ground-truth developer action.
  \item \textbf{Model-Based Evaluation}: ``LLM-as-a-Judge'' autoraters that assess semantic correctness and helpfulness.
  \item \textbf{Execution Metrics}: For tasks like automated repair, we verify correctness via compilation and test execution.
\end{enumerate}

This continuous evaluation protocol provides robust signals, enabling precise validation of model convergence and selection of optimal checkpoints.

\subsection{Data Mixture and Ablation Strategy}
In order to determine the final mixtures, we run several ablations with models of smaller sizes and different weights assigned to different categories or different datasets in the mixtures. We then observe the impact on evaluations to identify the best mixture weights. Running these ablations with smaller models enables us to run ablations quickly and efficiently. Once we have the final mixtures derived from these ablations, we then scale our training to larger models.

\section{Post-training}
\label{sec:posttraining}

Following mid-training, the model undergoes a post-training phase to optimize its utility for downstream applications. This phase utilizes Gemini's post-training methodologies and curated dataset mixtures, augmented with high-quality internal datasets. These proprietary datasets are derived from crowdsourced validation from employee volunteers or from anonymized logs of  interactions with internal products and tools, undergoing refinement for quality through expert human review.

The sources used to generate these internal datasets (at the time of this writing) include:\footnote{The analysis of synthetic or agent-generated trajectory data in SFT or RL falls outside the scope of this article.}

\begin{figure}
  \centering
  \includegraphics[width=\linewidth]{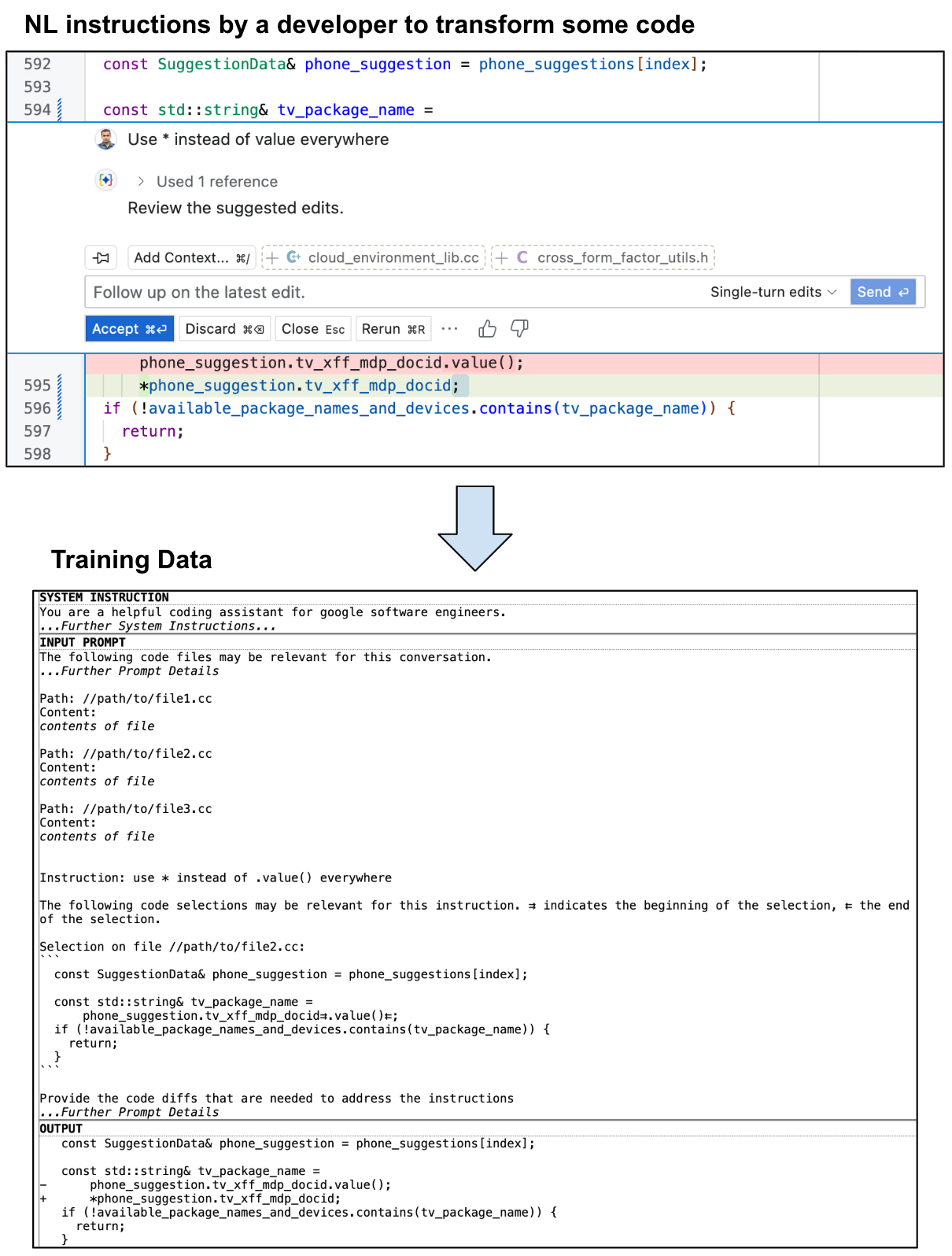}
  \caption{Illustration of how a logged interaction between the developer and the NL based code transformation feature in the IDE can be leveraged to generate training data.\label{fig:logged_transform}}
\end{figure}

\begin{figure}
  \centering
  \includegraphics[width=\linewidth]{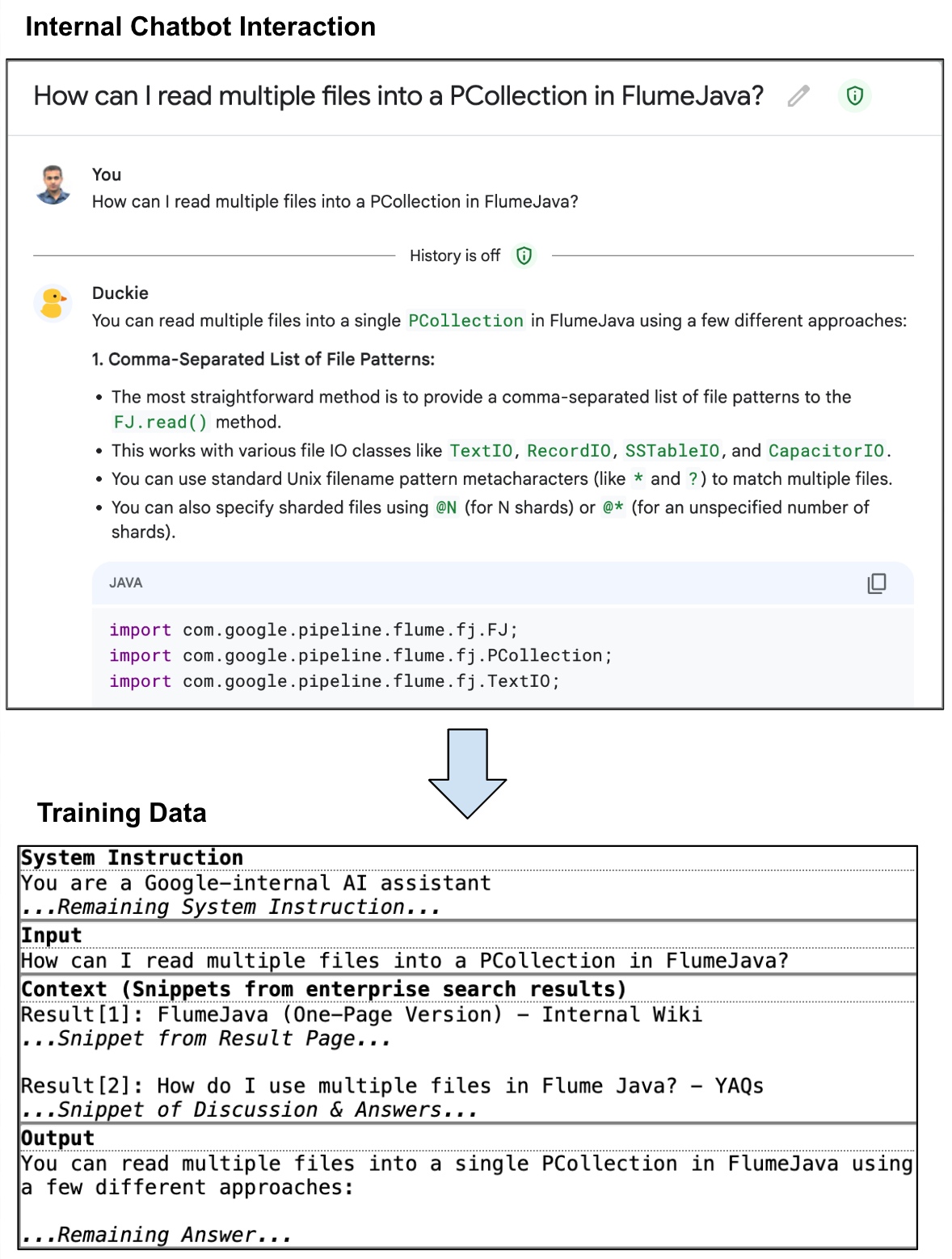}
  \caption{Illustration of how a logged interaction between the developer and the internal developer assistance chatbot is converted to a post-training example.\label{fig:logged_qa}}
\end{figure}

\begin{itemize}
  \item \textbf{Code Transformation ~\cite{tabachnyk2022ml}}: Logs capturing natural language instructions for code modifications within the Cider IDE, paired with model-generated outputs and subsequent user feedback (acceptance, rejection, or rewriting), providing a measure of output quality. This dataset represents many different types of interactions of the developers with the code transformation capabilities in the IDE, including refactorings of the code, implementing methods when the docstrings and signatures have been created, generating tests for already written code or adding logic described in an NL prompt to the existing code (see Figure~\ref{fig:logged_transform}).

  \item \textbf{Migrations}: A carefully compiled collection of high-quality prompts and corresponding model outputs generated through usage of an internal code migration tool, facilitating the transition of the codebase from legacy to contemporary technologies.

  \item \textbf{NL2SQL}: A comprehensive corpus of SQL queries paired with their corresponding natural language descriptions, crowd-sourced from internal software engineers.

  \item \textbf{Q\&A}: A refined set of high-quality question-and-answer pairs extracted from the interaction logs of internal chatbot systems and other knowledge retrieval platforms (see Figure~\ref{fig:logged_qa}).

\end{itemize}

These datasets undergo rigorous manual curation to ensure high data quality. A representative subset of examples from each dataset is held out for evaluation purposes. Subsequently, these refined datasets are integrated with Gemini post-training instruction tuning mixtures and this combined mixture is used during the Supervised Fine Tuning phase of our post-training.

\begin{figure}
  \centering
  \includegraphics[width=\linewidth]{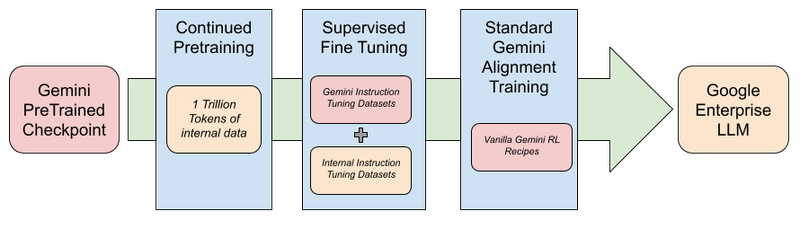}
  \caption{The E2E training process}
\end{figure}

\section{Results}
\label{sec:results}

In the following, we present some lessons learned from applying this kind of model
customization in an enterprise setting using Gemini for Google (\textit{GfG}) within the Google ecosystem.
We designed our evaluation to answer the following primary questions, and finally
highlight some additional lessons:

\begin{enumerate}[label=\textbf{Q\arabic*:}, leftmargin=2em]
  \item \textbf{Model Proficiency.} How does the finetuned \textit{GfG} model compare to
    the Gemini baseline on standard public benchmarks?

  \item \textbf{Industrial Utility.} What is the impact of \textit{GfG} on developer
    productivity and code adoption rates in a live enterprise environment?

  \item \textbf{Developer Assistance.} To what extent can an enterprise-tuned LLM enhance software engineering productivity by automating high-volume maintenance, optimizing resource efficiency, and augmenting interactive developer workflows?

  \item \textbf{Mitigating Catastrophic Forgetting.} What kind of regressions do we see when creating a specialized enterprise LLM? How do we mitigate such regressions, if discovered?

  \item \textbf{Lessons from Data.} What are the insights and drawbacks that we discovered during the curation and scaling of the trillion-token proprietary dataset?
\end{enumerate}

\textit{Model Setup:} In the evaluations below we compare the Gemini 2.5 Pro model with the equivalent Gemini for Google version. While newer model iterations (e.g., Gemini 3) exist, we employ Gemini 2.5 Pro to maintain consistency with our extensive internal benchmarking datasets, which were most complete for the 2.5 architecture during the data collection period.

\subsection{Offline Evaluation (Q1)}
Assessing model proficiency requires a combination of external and internal benchmarks covering both coding tasks and general reasoning. Our internal benchmarks assess the model's ability to execute tasks representative of a Google Software Engineer's daily workflow, while the external evals are chosen for their reliability and established use in evaluating frontier models. For example, we relied on SWE-bench Verified \cite{chowdhury2024swebenchverified} for its focus on resolving real-world GitHub issues, Aider \cite{aider_leaderboard} for multi-turn editing, and LiveCodeBench \cite{jain2024livecodebenchholisticcontaminationfree} for evaluation of broader coding tasks beyond code generation.

\begin{table}
  \centering
  \begin{tabular}{lc}
    \toprule
    \textbf{Evaluation Metric} & \textbf{Relative Performance} \\
    \midrule
    \multicolumn{2}{c}{\textbf{External Benchmarks}} \\
    \midrule
    SWE-bench Verified & 1.05 \\
    Aider & \\
    \hspace{1em} Polyglot (Whole File) & 1.04 \\
    \hspace{1em} Polyglot (Edit Block) & 1.13 \\
    \hspace{1em} Polyglot (Unified Diff) & 0.94 \\
    \hspace{1em} Polyglot Multiturn (Edit Block) & 1.01 \\
    LiveCodebench & \\
    \hspace{1em} Execution & 1.01 \\
    \hspace{1em} Unit Test & 0.97 \\
    \multicolumn{2}{c}{\textbf{Internal Benchmarks}} \\
    \midrule
    Google SWE Tasks  & 1.05 \\
    Google Code Transformations & 1.04 \\
    \bottomrule
  \end{tabular}
  \caption{Performance of the \textit{GfG} checkpoint compared to Gemini 2.5 Pro model on some external and internal coding benchmarks. Values $> 1$ indicate \textit{GfG} superiority.}
  \label{tab:gemini-for-google-performance}
\end{table}

As seen in Table~\ref{tab:gemini-for-google-performance}, \textit{GfG} generally outperforms the baseline Gemini checkpoint on the majority of the coding benchmarks. The model performs marginally better on the Aider benchmarks, notably with a 13\% improvement on the Edit Block format that requires ``Search and Replace'' blocks. \textit{GfG}'s proficiency comes from the ``critique-and-refine'' examples described in \S\ref{sec:data_curation} where the model is trained to handle the block-like nature of code reviews, where comments are strictly associated with specific code snippets. While \textit{GfG} shows slight regression on the unified diff (udiff) format likely due to the strict line-prefix required for exact context matching, it remains competitive on whole file and multi-turn editing.

Despite the trillion-token injection, Table~\ref{tab:non-coding-performance} confirms that cognitive reasoning and multimodal capabilities are preserved. \textit{GfG} maintains or improves over baseline performance on ARC AGI \cite{chollet2025arcprize2024technical} and GradQA which respectively measure the ability to solve abstract puzzles and graduate-level reasoning. To verify multimodal stability, we used MMMU-Pro \cite{yue2025mmmuprorobustmultidisciplinemultimodal} which tests reasoning across both visual and textual formats. while exhibiting marginal variances in expert-level domain assessments like Humanity's Last Exam (HLE) \cite{phan2025_hle_humanity_last_exam_paper} and IMO AnswerBench \cite{luong2025robustmathematicalreasoning}.

\begin{table}[t]
  \centering
  \begin{tabular}{lc}
    \toprule
    \textbf{Evaluation Metric} & \textbf{Relative Performance} \\
    \midrule
    \textbf{Hidden Math - Hard Subset} & \\
    \hspace{1em} avg@4 & 0.99 \\
    Humanity's Last Exam & 0.84 \\
    \midrule
    \textbf{IMO AnswerBench V1} & \\
    \hspace{1em} pass@1 & 0.94 \\
    \midrule
    ARC AGI & 1.04 \\
    GradQA & 1.03 \\
    MMMU - Pro \faCamera & 0.96 \\
    \bottomrule
  \end{tabular}
  \caption{Relative performance of \textit{GfG} compared to the baseline Gemini on non-coding benchmarks to monitor for regression. Despite massive domain-specific training, both the fundamental intelligence and multimodal reasoning of the base model were preserved throughout the mid-trainign intervention and subsequent post-training.}
  \label{tab:non-coding-performance}
\end{table}

\subsection{Online Evaluation (Q2)}
The quantitative and qualitative impact of the fine-tuning can also be measured through an A/B test. We ran an A/B test comparing baseline Gemini with the fine-tuned \textit{GfG} model to measure the production impact. The following section lists the details of our observations:

\subsubsection{\textbf{Experimental Setup.}}
To rigorously assess the performance of \textit{GfG} against the baseline model, we conducted a blind, randomized A/B experiment in September 2025. We evaluated the model's performance across 29,000 active users by deploying Gemini for Google into the critical development workflows and IDEs that power engineering across Google. Agentic coding development in IDE's was the major use case. The agentic coding development paradigm empowers the model to autonomously plan, edit across multiple files, and iteratively self-correct using test feedback, effectively elevating the developer's role from manual implementation to high-level orchestration. Within this framework, prompting becomes a critical interface: rather than simple queries, prompts serve as strategic specifications that encode the intent and constraints required for the reasoning engine to navigate complex, multi-step engineering tasks with minimal oversight. The model served as the reasoning engine for an integrated agentic coding workflow, facilitating real-time developer assistance.

To eliminate selection bias, users were blindly assigned to either the treatment group (\textit{GfG}) or the control group (Gemini) for the duration of the experiment. The experiment compared Pro models (based on Gemini 2.5 Pro) and Flash models, with Pro models accounting for 97\% of the total usage. The analysis focused on data collected over seven full working days to ensure statistical validity. A power analysis confirmed that this duration provided sufficient statistical power to detect significance across the majority of key metrics.

\subsubsection{\textbf{Quantitative Results}}
GfG demonstrated statistically significant improvements over the baseline across all critical productivity and quality metrics (see Table~\ref{tab:industrial_impact}):
\begin{table}
  \centering
  \caption{Qualitative comparison of \textit{GfG} vs Gemini 2.5 Pro for key online performance metrics. These results highlight improvement across the board for \textit{GfG}.}
  \label{tab:industrial_impact}
  \begin{tabular}{llc}
    \toprule
    \textbf{Dimension} & \textbf{Key Metric} & \textbf{Relative Change} \\ \midrule
    \multirow{2}{*}{Efficiency} & Conversational Iterations & $-23.1\%$ \\
    & Turn Latency & $-8.9\%$ \\ \midrule
    \multirow{2}{*}{Quality} & Hunk Acceptance Rate & $+4.5\%$ \\
    & Code Line Acceptance & $+4.5\%$ \\ \midrule
    \multirow{2}{*}{Impact} & Code Submission Rate & $+11\text{--}14\%$ \\
    & Code Survival Rate & $+16.8\%$ \\ \bottomrule
  \end{tabular}
\end{table}

\begin{itemize}
  \item \textit{Efficiency and Latency}: \textit{GfG} significantly reduced the cognitive and temporal load on developers. We observed a 23.14\% reduction in the mean number of iterations per turn, indicating that Gemini for Google required fewer conversational turns to satisfy user intent. Additionally, the mean end-to-end turn duration decreased by 8.85\%.
  \item \textit{Code Acceptance}: Code quality, measured by user acceptance, showed clear gains. The hunk acceptance rate (the percentage of contiguous model-generated blocks, also called hunks, accepted by the developer) was 4.49\% higher for \textit{GfG}, while the code line acceptance rate experienced a parallel growth of 4.45\%.
  \item \textit{Production Impact and Code Survival}: \textit{GfG} drove higher engagement and long-term value. There was an 11---14\% increase in submitted code changes utilizing the agent in the treatment group. Furthermore, code generated by \textit{GfG} exhibited better persistence; the {\em survival rate} of accepted code in final submissions was 16.80\% higher compared to the baseline. In an ecosystem where over 75\% of all code is already written or drafted by AI, submission and survival rates better capture actual engineering value and true utility. \textit{GfG}'s superior performance on both these fronts shows that it effectively bridges the gap between raw AI generation and enduring, production-grade software.
\end{itemize}

\subsubsection{\textbf{Qualitative Observations}} User feedback revealed a gap between models that highlight the benefits of domain specialization in \textit{GfG}:
\begin{itemize}
  \item \textit{Perceived Quality Drop}: Users in the control group reported that the baseline model felt less capable and represented a  quality regression.
  \item\textit{Domain-Specific Struggles}: The baseline model required significantly more turns to correctly implement internal build rules and imports compared to \textit{GfG}.
  \item\textit{User Friction}: The performance gap was significant enough that users sought workarounds to bypass the experiment and restore access to \textit{GfG}.
\end{itemize}

\subsection{Case Studies in Industrial Application (Q3)}
We evaluate the model's deployment across three distinct modes of operation: large-scale automated refactoring (migrations), autonomous code optimization, and automatic paste fixing.

\begin{enumerate}[leftmargin=*, nosep]
  \item \textbf{Migrations} ~\cite{nikolov2025googleusingaiinternal}

    Google successfully employed this model to migrate substantial, decade-old codebases for certain product areas away from older systems.
    \subsubsection{Case Study A: Infrastructure Modernization (Ads product area)}
    A notable instance was the migration within Ads product area, transitioning from 32-bit numerical IDs for entities like users and campaigns to 64-bit integers to avoid overflow errors. It is estimated that the total time was reduced by 50\% compared to manual methods, with AI fully authoring 80\% of the code changes in submitted changelists.

    \subsubsection{Case Study B: Framework Migration (JUnit)}
    This model was used to migrate a substantial set of legacy JUnit3 test files to the modern JUnit4 framework. This automated approach successfully migrated 5,359 files (over 149,000 lines of code) in 3 months, with approximately 87\% \cite{tabachnyk2022ml} of the AI-generated code committed without human modification, the main bottleneck being the human review process.

  \item \textbf{Code Efficiency} ~\cite{lin2025ecollmdrivenefficientcode}
    Another application of this model is ECO (Efficient Code Optimizer). ECO detects performance anti-patterns in the warehouse-scale codebase and utilizes the model to synthesize optimization patches.
    \begin{itemize}
      \item \textit{Pipeline}: The model acts as the core reasoning engine, taking profiling data and source code as input to generate refactors, which are then verified and submitted for human review.
      \item \textit{Scale of Impact}: The system has deployed over 6,400 commits (>25,000 lines of changed production code) with a >99.5\% production success rate.
      \item \textit{Hardware Savings}: These optimizations translate to substantial hardware efficiency, saving an estimated 500,000 normalized CPU cores per quarter.
    \end{itemize}

    Beyond the described use cases above, this model trained on a wide range of internal enterprise datasets has seen widespread adoption across Google's internal development infrastructure. Its applications include, but are not limited to:
    \begin{itemize}
      \item \textit{Automated Unit Test Generation}: Reducing the friction of increasing test coverage.
      \item \textit{Production Trace Analysis}: Identifying bottlenecks by correlating server traces with source code.
      \item \textit{Conversational Assistance}: An internal chat application for architectural Q\&A.
    \end{itemize}

  \item \textbf{In-IDE Assistance}  ~\cite{nguyen2026smartpasteautomaticallyfixing}
  
    ``Smart Paste'' is a production IDE feature that utilizes the model to intelligently adapt copied clipboard code content to its destination context. This tool addresses the frequent need for manual ``fix-up'' edits—such as reconciling imports, variable names, and formatting—immediately after a paste action. Utilizing \textit{GfG} was essential to handle proprietary libraries and meet strict millisecond-latency requirements for an in-flow experience.

    \begin{itemize}
      \item \textit{Adoption and Impact}: The feature has seen widespread adoption, utilized daily by tens of thousands of developers with a 45\% acceptance rate.
      \item \textit{Efficiency}: Each accepted suggestion saves an average of 22 keystrokes (capturing deletions and replacements), and collectively, code generated by Smart Paste now accounts for over 1\% of all code produced.
      \item \textit{Code Survival}: The model's suggestions demonstrate high utility, with approximately 58\% of the generated characters remaining after 30 minutes.
      \item \textit{Emergent Workflows}: Beyond standard adaptation, the model's capabilities enabled developers to use copy-paste as a tool for lightweight repetitive refactoring (``chaining'') and context-aware cross-language translation.
    \end{itemize}

    ``Transform Code'' is an implementation of inline chat that allows for a natural language instruction to modify a highlighted section of code. \textit{GfG} provides the basis for contextual awareness of internal code.

    \begin{itemize}
      \item \textit{Productivity}: Early adopters of inline chat showed a 15\% increase in throughput over the control.
      \item \textit{Acceptance}: 35,000 accepted suggestions per week, with a 70\% acceptance rate.
    \end{itemize}

\end{enumerate}

\subsection{Mitigating Catastrophic Forgetting (Q4)}
A core challenge in mid-training intervention is retaining the general reasoning capabilities of the base model while injecting domain knowledge.

\textbf{Problem}: Our initial methodology involved starting \textit{GfG}'s continued pre-training from  the 100\% Gemini pretrained checkpoint. While the model excelled at internal brownfield tasks, its general reasoning and foundational proficiency on standardized tasks degraded due to the distributional shift between internal data and the general web corpus.

\textbf{Solution}: We adopted two strategies:
\begin{enumerate}
  \item \textit{Data Replay}: We mixed some of the original pre-training data (parent model data) into the enterprise mixture. This anchors the model's weights, preserving general capabilities.

  \item \textit{Mid-training intervention}:  Instead of extending training from the final checkpoint , we branch off from an earlier checkpoint (e.g., before the final convergence phase of the base model).
\end{enumerate}

\begin{table}[h]
  \small
  \centering
  \begin{tabularx}{\columnwidth}{Xccc} 
    \toprule
    \textbf{Model Variant} & 
    \textbf{\begin{tabular}[c]{@{}c@{}}Data Science\\ Workflows\end{tabular}} & 
    \textbf{\begin{tabular}[c]{@{}c@{}}NL Code\\ Synthesis\end{tabular}} & 
    \textbf{\begin{tabular}[c]{@{}c@{}}Competitive\\ Coding\end{tabular}} \\ \midrule
    Gemini & 45.8 & 74.6 & 16.6 \\
    \textit{GfG} with continued pre-training & 11.9 & 36.3 & 1.3 \\
    \textbf{\textit{GfG} mid-training} & \textbf{39.0} & \textbf{70.3} & \textbf{12.0} \\ \bottomrule
  \end{tabularx}
  \caption{Performance comparison showing the benefits of data replay and mid-training in mitigating catastrophic forgetting on different reasoning and general coding tasks. These benchmarks evaluate proficiency in data science problem-solving, generating code from natural language intent, and solving greenfield programming problems. Results are reported as percentage scores (pass@1).}
  \label{tab:model_performance}
\end{table}

Additional changes were also needed for adapting the mixture of expert model architecture for specialized use cases. Mixture-of-Experts (MoE) architectures are typically designed as generalists. We observed that standard router load-balancing mechanisms failed during domain-specific fine-tuning. During training, evaluation metrics saturated at approximately 45\% of the run. Investigation revealed significant token dropping, as the router disproportionately favored specific experts for the highly homogenous internal data. Hence, we needed to customize the expert routing technique to prevent expert collapse.

\section{Lessons Learned (Q5)}
\label{sec:lessons}
Through the development and deployment of our software engineering models, we identified critical challenges in data curation, alignment, and evaluation. We distill our experience into the following three lessons:

\begin{enumerate}
  \item \textit{Data Composition}: The maxim ``more data is better'' does not always hold --- pre-training distributions can establish priors that conflict with post-training.

    \textit{Observation}: We included a ReAct-style dataset in pre-training intended to improve reasoning. However, this dataset interacted poorly with the post-training phase, degrading the model's function-calling behaviors.

    \textit{Insight}: This highlights negative transfer, where learning a specific formatting style (e.g., verbose reasoning traces) interferes with the precise syntax required for tool use. Careful data ablation is required to balance ``learning to code'' vs. "learning to use tools."

  \item \textit{Schema Alignment Minimizing Distributional Shift.} Fighting the base model's data format is computationally expensive and architecturally risky.

    \textit{Observation}:  Early attempts to restructure data into custom internal formats led to regression issues during post-training.

    \textit{Takeaway}: Aligning internal data to match the pre-training format of the base model (Gemini) proved essential. It minimizes the "syntax shock" to the model and simplifies the transition between public and proprietary capabilities.

  \item \textit{The Evaluation Gap Beyond Static Benchmarks.}
    Evaluating generative models for enterprise engineering remains an unsolved problem. Standard ``pass@k'' metrics on unit tests do not capture the nuance of maintaining legacy code. We found that execution-based metrics often failed to fully capture developer satisfaction. Consequently, we augmented our evaluation strategies with ``Autorater'' assessments (LLM judges) and live production telemetry (acceptance rates) in addition to static benchmarks.
\end{enumerate}

\section{Conclusion}
We introduced Gemini for Google, a specialized adaptation of Gemini tailored for Google's internal software engineering ecosystem. By fine-tuning on a trillion-token proprietary corpus and employing novel strategies like mid-training intervention we successfully mitigated catastrophic forgetting while capturing complex enterprise semantics. A large-scale blind A/B study across 29,000 developers demonstrated that Gemini for Google significantly outperformed baselines. Ultimately, this work provides a replicable blueprint for organizations to leverage their own engineering data for full-stack model adaptation, from data curation to downstream application deployment.

\section*{Data Availability Statement}
The datasets, codebases, and infrastructure utilized in this study---including internal IDE telemetry, code review logs, and proprietary training mixtures---are confidential and proprietary. Due to stringent data governance, corporate security policies, and privacy protocols, these artifacts cannot be made publicly available. We have provided comprehensive methodological details in the paper to ensure the principles of our approach can be understood and replicated by the community on alternative datasets.

\section*{Acknowledgements}
We thank our colleagues at Google, particularly Oriol Vinyals, Niranjan Tulpule, Madhura Dudhgaonkar, Leonard Berrada, Dawn Chen, Hannah Lin, Patrick Musau, Amita Gondi, Arman Hasanzadeh, Dan Zheng, Deniz Alt\i nb\"uken, Fred Lewis, Hannah Lin, Maxim Tabachnyk, Pat Rondon, Sandeep Katragadda, Sergei Shmulyian, Stephanie Tang, Stoyan Nikolov, Kashmira Phalak, Michael Golahi, Harshit Vadodaria, Sandeep Katragadda, Jerry Peng, Martin Dixon, and David Lo.

\bibliography{arXiv/gemini-for-google}

\end{document}